\documentclass[sigconf,natbib=false]{acmart}

\AtBeginDocument{%
  }

\setcopyright{acmlicensed}
\copyrightyear{2024}
\acmYear{2024}
\acmConference[IDE '24]{2024 First IDE Workshop}{April 20, 2024}{Lisbon, Portugal}
\acmBooktitle{2024 First IDE Workshop (IDE '24), April 20, 2024, Lisbon, Portugal}
\acmISBN{979-8-4007-0580-9/24/04}
\acmDOI{10.1145/3643796.3648457}

\acmSubmissionID{ICSE-WORKSHOP-IDE-17}

\RequirePackage[
  datamodel=acmdatamodel,
  style=acmnumeric,
  ]{biblatex}

\begin{document}

\title{Software development in the age of LLMs and XR}

\author{Jesus M. Gonzalez-Barahona}
\email{jesus.gonzalez.barahona@urjc.es}
\orcid{0000-0001-9682-460X}
\affiliation{%
  \institution{Universidad Rey Juan Carlos}
  \city{Fuenlabrada}
  \country{Spain}
}

\renewcommand{\shortauthors}{Gonzalez-Barahona}

\begin{abstract}
  Let's imagine that in a few years generative AI has changed software development dramatically, taking charge of most of the programming tasks. Let's also assume that extended reality devices became ubiquitous, being the preferred interface for interacting with computers. This paper proposes how this situation would impact IDEs, by exploring how the development process would be affected, and analyzing which tools would be needed for supporting developers.
\end{abstract}



\received{21 January 2024}

\maketitle

\section{Hypothesis}

During the next few years, software development will change. Generative AI (artificial intelligence), mainly via machine learning LLMs (large language models) is already being used for assisting programming tasks. From pair programming with a generative AI agent to automatic generation of code, explanation of it, or generation of testsuites, new ways of coding are being explored worldwide. If the promises of this new technology deliver, we can expect radical changes in how software is developed and maintained, with a much more intense participation of generative AI in the process.

Also in a few years, the main interface to interact with computers may change. Lighter, more powerful yet less energy-consuming virtual and augmented reality appliances are expected to be developed, leading to devices with the form factor of regular glasses, the computing power of a desktop, and the connectivity of a mobile phone. If this trend materializes, the preferred mean for interacting with computers could be XR (extended reality).

This paper is not about how these two tendencies may become real. It is about what would happen if they become real. Our focus here is to explore what would happen in that world, and in particular, how would software be developed and maintained, and which kind of tools would would software developers need. Those tools would be the IDE (integrated development environment) of the future.

\section{Generative AI}

  The idea of automatic generation of software is not new. Early compilers were in many cases framed as ``automatic code generators'', since high level languages were a significant advance in abstraction when compared with assembler. More recently, model-driven development~\cite{pastor200:mdd} has been exploring how to describe requirements for software systems, and then automatically generate code meeting those requirements. Requirements are expressed in domain specification languages, tailored to experts of the concerned domains, and code is produced automatically from them, maybe providing hooks for specific parts of the programs to be coded ``manually'', if the need arose. However, this and other similar approaches have not been successful except for some specific, relatively small domains.

  The emergence of generative AI has changed the landscape, showing it is capable of generating many different kinds of content, including source code~\cite{cao2023:ai-generated-content, gozalobrizuela2023:review-generative-ai}. In the particular case of code, LLMs are already being used or researched for many software engineering tasks~\cite{hou2023:llms-software-engineering, ebert2023:generative}, from automatic code generation from natural language descriptions of requirements or code summarization and explanation~\cite{chen2021:llms-code}, to test generation~\cite{wang2023:llms-software-testing} or bug fixing~\cite{joshi2023:llms-repair}. Although it is showing some shortcomings, such as the production of  plausible yet wrong code, of low quality code, or of code with security flaws~\cite{sun2022:explainability-gai, gupta2023:threatgpt}, the approach seems very promising.

  It is still difficult to predict how this new technology will be used in the future, but it could happen that more and more, coding becomes a task performed by generative AI agents, while developers focus on expressing what programs should do in natural language. Of course, there will be niches, may be many of them, where programming will still be completely done by human developers, but even in these cases they will have a lot of support by generative AI.

  \section{Extended reality}
  
As in most other application domains, the standard user interface for software development has been based on 2D screens, mainly showing code as text. If XR devices become the usual interface, interfaces could change dramatically, and we could explore how to take advantage of the natural interactions of humans with their surrounding world, based on movement and spatial perception~\cite{elliott15:_virtual_realit_softw_engin}.

Many engineering disciplines already use 3D-based devices. In recent times, they are starting to take advantage of XR to accommodate better for 3D-based digital interactions, with concepts such as ``digital twins'' (virtual representations of physical entities) in XR to simulate the behavior of complex systems~\cite{el_saddik_digital_twins:2018}. Software engineering could take advantage of XR too, overcoming the limitations of 2D screen-based interactions. This was already recognized in 1999~\cite{knight_comprehension_1999}, when virtual environments were presented as convenient to overcome the cognitive overload caused by the many visualizations needed to understand the many aspects of software development.

Already at that time there were proposals for using software visualization to facilitate comprehension and reasoning about complex systems~\cite{harel1990:statemate}. Since then, software visualization has been an active field of research. In particular, 3D-based visualization has been explored for more than 20 years~\cite{teyseyre2008:3d_software_visualization,chotisarn2020:review_software_visualization}. However, the limitations of the substrate in which 3D metaphors were implemented (2D screens) limited their possibilities. XR interfaces allow for more information available simultaneously, which is good for representing the many aspects of software, and to interact with 3D objects in more intuitive ways.

  \section{The IDEs of the future}

  In case this two trends (generative AI in charge of coding, and XR as user interface) become mainstream, which kind of software support will we need for developing software? Of course, this will depend on how we combine our abilities with those of generative AI, and on how we use the affordances offered by XR interfaces. But whatever the answer, two main problems will have to be solved: (1) how to produce reliable pieces of code, following some specification, and (2) how to combine them in large, complex systems that work. Of the many different possibilities, the following proposal could be used as a starting point for a more nuanced discussion.

  \subsection{Producing reliable code}
  
  Considering the first problem, we can observe how that problem is addressed in current software development processes. In particular, we propose as a model how reliability and code quality is improved in many large FOSS (free, open source software) projects. In them, changes to the source code follow a well established process before being accepted: developers fix bugs, improve the software, or add new functionality by producing patches, which are thoughtfully reviewed, discussed and then accepted into the code base, in a cycle that may require several iterations over the proposed patch. In addition, it is more and more usual to have detailed testing policies, which usually require tests to be included in the patch, and regression and integration tests to be passed before considering acceptance~\cite{napolean2020:oss-development-process}. This causes a natural division of tasks in the project, with most developers devoted to writing changes to the code, and tests for it, while a small number of more experienced developers tend to devote most of their time to review. Reviewers may be assisted by automatic tools for code analysis and testing, which let them focus on changes that are worth checking.

  We could envision an evolution of this model with more and more generative AI in the loop. Code changes could be produced more and more automatically, with humans focusing on describing what the code should do, and reviewing the produced code. Tests could also be built automatically, and other tools could help to ensure that the code meets some quality standards. In the end, humans could just explain to generative AI agents how they want the code to perform, so that agents produce the code and comprehensive sets of tests. Humans would later check that the testing cases are complete enough, and that they pass before considering accepting the patch. This model has already being tried in part~\cite{kshitij2023:llmcodingagent}.

  \subsection{Composing pieces of code}
  
  Composing small pieces of code to build large-scale, real-world systems, and understanding how they work together, is challenging. Despite the many software engineering techniques devoted to address this problem, in real cases we still we rely on the knowledge of experts familiar with the system. Even for just representing the system in whole we lack good approaches that let us change scale and help us to understand how it works at different levels.

  Extended reality may allow for evolving towards new ways of approaching the software more natural to our cognitive capacities. As other disciplines use 3D models intensively, we could learn to approach to software via similar models. We could try metaphors for software comprehension better suited to our capabilities than those possible with screens of text. This could help us to better deal with software at an architectural level, as a complex set of modules interacting with each other. Generative AI can already help in this comprehension, by providing summarization and by generating modules that use APIs from other modules. But the general architecture of the system, the interactions between its modules, etc. could be for a long time a domain for AI-assisted humans.

  \section{One step further}
  \label{sec:example}

    Combining both approaches sketched above, the IDE of the future could be a 3D graphical interface, framed around some metaphor that allows us to better comprehend software complexity at an architectural scale, and providing means for interacting with generative AI agents that would build, repair and improve code at the source code scale.

  A natural step further could enable the production and maintenance of complex software systems without humans having to interact with source code at all: we can apply the previous discussion to zero-code development. Generative models are already proving how they can produce, with some reliability, small code fragments following specifications. We can explore how to improve the process by using tests, and how to let domain experts drive the production of the code without actually interacting with the code. Then, for combining these code fragments into complete systems, we can explore how to use XR-based graphical interfaces.

  The production of small pieces of code, such as functions, could start with a domain expert specifying, in natural language, how the code should perform. Depending on the capabilities of the model, that specification could also include hints for the implementation, such as which algorithm to use. Then, the process for producing the code is based in two incremental loops, both assisted by generative models:
  
  \begin{itemize}
  \item Test production. The specification produced by the expert is used as input to a generative model, which is instructed to produce a collection of tests for it. Since the collection could be incomplete or include erroneous tests, the generative model will be asked to produce an explanation of the tests, including the input and the expected output for each of them. The expert will review that explanation, asking for completing or changes to the testsuite, which the generative model will produce. The process will continue, with the expert refining the tests until they are considered valid.
  \item Code production. Once the testsuite is adequate, the generative model is asked to produce the piece of code using the testsuite and the specification as input, and a second refinement cycle starts. This one is completely automated: the testsuite is run on the produced code, and if any test fails, the output of the execution, along with the specification and the testsuite is fed the generative model, until all tests pass.
  \end{itemize}

  This way, the final version of the code will be compliant with the testsuite. If the expert has training on how to build good testsuites, and good prompts for the specification, it is very likely that the implementation is correct. Other tools, such as static analysis tools, could also be included in the second loop, to ensure that the code has certain properties. Tools to analyze code quality could be used to select the best implementation among a certain number of candidates produced in the code production tool, too. 

  But functions or small pieces of code are only a part of the story. To build complete systems, those pieces have to be interconnected. This can be done with a graphical user interface, using data flow, code blocks, or some other mechanism that allows for composition of the pieces of code produced with the previous process. Domain experts could also design an integration testsuite with the help of a generative model, and debug and adjust the system using a process similar to the one used for small pieces of code.

  A similar process could be envisioned for finding the faulty code causing some malfunction, asking generative models to summarize the behaviors of different components based on logs, debug traces and code, with the expert interpreting those summaries, asking for additional probes in the code or new tests, until the problem is found and fixed.
  
  For working this way, domain experts would need new supporting software (their IDE), which they would use with a combination of speech, text (for reading details) and 3D models. Those IDEs would allow them to create large-scale software systems without reading or writing source code, the same way current programmers develop executable code without reading or writing executable code. The IDE could be extended to monitor the running of the produced system, once deployed, with the help of generative models to interpret the logs of the working of the system.
  
  \section{How to get there}

  To reach to that kind of IDEs, we still need to follow a large research and development agenda. Some of the main lines that should be explored with detail are:

  \begin{itemize}
  \item Integration with generative AI. Even when chatting with generative AI agents and letting them complete code are today the preferred ways of interacting with them, we can expect that the future will also include many actions initiated by AI agents themselves. For example, they could react to malfunctions or bug reports by finding bugs and proposing how to fix them, or being in a continuous quest for improving aspects of the software with or without human intervention. The interaction with agents should likely include not only talking to them, but also discussing with them about diagrams, examples, etc, in ways that are still to be found.
  \item Collaboration with remote agents and humans. Software development is already a collaborative process. In the future, collaboration with remote people and agents will become commonplace. Therefore, we will need to have this collaboration thoughtfully integrated in our development environments.
  \item Metaphors of how to represent software in extended reality scenes. Software, not having a real-world physical manifestation, also does not have a preferred representation in XR. This gives plenty of opportunities for finding which metaphors could be better exploited by our cognitive capabilities to better comprehend and interact with software complexity.
  \item Interactions with software metaphors in XR. Even when the keyboard may still be an input method in XR, it is to be expected that other means, such as speech and gestures, could also be used.
  \item New skills and training for producing software. For example, if some software is to be developed as shown in the zero-code example, domain experts could be trained on designing specifications and good descriptions of testsuites, in a way that can be used by generative models. We will have to design new learning paths for these ``zero-code developers'', and adapt the IDEs to that training.
  \end{itemize}

  For sure there are many other challenges. But just from this short list, it is clear that in case the scenario described in this paper becomes real, the change to the IDEs needed in it will be disruptive.

  \section{Conclusions}

  In a world where generative models are the preferred way of producing code, and XR is the main user interface, IDEs will still be needed. They will be used to interact with developers in new ways, adapted to their training, and to the new software production processes. We expect that interaction to be based on speech and graphical interfaces, so that IDEs could seem to disappear behind a high-level interface, but they will still be needed to support the needs of developers behind the curtains.

  The code generation process described in the example in Section~\ref{sec:example} is being explored with a simple proof of concept, still under development\footnote{\url{https://gitlab.com/jgbarah/currante}}.
  
  Please, consider all this text just as food for thought. Of course, the future is difficult to predict, and things will evolve in unpredictable ways.

\begin{acks}
  This study has been funded in part by the Spanish Government, via project Dependentium, PID2022-139551NB-I00.
\end{acks}

\printbibliography

\end{document}